\documentclass[preprint,showpacs,amsmath,amssymb,pra,supercriptaddress,floatfix]{revtex4}
\usepackage{graphicx}
\usepackage{dcolumn}
\usepackage{bm}

\begin{document}

\title{Atomic hyperfine resonances in a magnetic quadrupole field}
\date{\today}
\pacs{32.60.+i,33.55.Be,32.10.Dk,33.80.Ps}
\author{Shahpoor Saeidian}
\email[]{s_saeid@physi.uni-heidelberg.de}
\affiliation{Physikalisches Institut, Universit\"at Heidelberg, Philosophenweg 12,
69120 Heidelberg, Germany}%
\author{Igor Lesanovsky}
\email[]{Igor.Lesanovsky@uibk.ac.at} \affiliation{Institut
Quantenoptik und Quanteninformation,Institut f\"{u}r Theoretische Physik, Universit\"at Innsbruck,
Technikerstrasse 21 a, A-6020 Innsbruck,
Austria}%
\author{Peter Schmelcher}
\email[]{Peter.Schmelcher@pci.uni-heidelberg.de}
\affiliation{Physikalisches Institut, Universit\"at Heidelberg,
Philosophenweg 12, 69120 Heidelberg, Germany}%
\affiliation{Theoretische Chemie, Institut f\"ur Physikalische Chemie,
Universit\"at Heidelberg, INF 229, 69120 Heidelberg, Germany}%

\date{\today}
\begin{abstract}\label{txt:abstract}
The quantum resonances of an atom possessing a single valence
electron which shows hyperfine interaction with the nucleus is
investigated in the presence of a three dimensional magnetic
quadrupole field. Particular emphasis is put on the study of the
interplay of the hyperfine and quadrupole forces. Analyzing the
underlying Hamiltonian a variety of symmetries are revealed
which give rise to a two-fold degeneracy of the resonance
energies. Our numerical approach employs the complex scaling method
and a Sturmian basis set. Several regimes and classes of short-lived and
long-lived resonances are identified. The energies and decay widths
of the resonances are characterized by their electronic and nuclear
spin properties.
\end{abstract}

\maketitle

\section{Introduction}

Ultracold atomic gases offer a wealth of
opportunities for studying quantum phenomena at mesoscopic and
macroscopic scales
\cite{Pethick02,Pitaevskii03,Ketterle99,Leggett01,Dalfovo99}. The majority of
experiments in the ultra-low temperature regime
have so far been performed with alkali atoms. Their ground-state
electronic structure is characterized by the fact that all electrons
but one occupy closed shells. The latter, the so-called
valence electron, is situated in the s-orbital of the outermost
shell. The coupling of the nuclear spin $I$ to the total angular momentum of the active electron via the hyperfine interaction causes a splitting of the electronic energy levels into several branches which can be characterized by the quantum number of the total spin $F$. Due to a vanishing orbital angular momentum this hyperfine interaction consist exclusively of the coupling of the electronic spin $S=\frac{1}{2}$, to the nuclear spin $I$.  This leads to
the two possibilities $F = I \pm S$.

Inhomogeneous magnetic field configurations represent a key
component for the control of the motion of cold atoms, specifically
for their trapping. The underlying Zeeman interaction occurs due to
coupling of the total magnetic moment of the atom to the external
magnetic field. Quantum states and particularly resonances of atoms
in inhomogeneous field configurations are therefore of
immediate interest to cold atomic physics in general and have been
investigated in the past by several groups. This includes the
quadrupole field \cite{Bergeman89,Lesanovsky05}, the wire trap
\cite{Berg-Soerensen96,Hau95,Burke96,Bluemel91} and the magnetic
guide and Ioffe trap
\cite{Hinds1,Hinds2,Potvliege01,Lesanovsky04,Bill06}. In these traps
we typically encounter quantum resonances with a certain lifetime
instead of stationary 'stable' quantum states. Very recently \cite{Saeidian06} a new class of short-lived
resonances possessing negative energies for the case of the 3d
magnetic quadrupole field was found. These resonances originate from a fundamental
symmetry of the underlying Hamiltonian.  In contrast to the positive
energy resonances, they are characterized by the fact that the
atomic magnetic moment is aligned antiparallel to the local
direction of the magnetic field, and their lifetimes decrease with
increasing total magnetic quantum number. 

In all of the above investigations it is assumed that the hyperfine
interaction is much stronger than the interaction with the external
inhomogeneous magnetic field and therefore the electronic and
nuclear angular momenta firstly provide a total angular momentum
which then interacts with the magnetic field. This physical picture
truly holds for (alkali) atoms in their electronic ground state and
macroscopic as well as microscopic (atom chip) gradient fields.  The atom is consequently treated as a point particle with the total angular momentum $F$.  In the present work we study the case for
which both interactions, the hyperfine and the field interaction,
have to be taken into account on equal level for the description of
the neutral atoms in the field. This case is primarily of principal
interest but is expected to describe the magnetized hyperfine
properties of systems with very small hyperfine interactions and/or
strong gradient fields, such as electronically excited atoms. In the
latter case an admixture of different hyperfine states $F$ due to
the field interaction has to be expected. In ref.
\cite{Gerbier06} the possibility to tune the hyperfine splitting by
dressing the electronic energy levels by a microwave has been
demonstrated. Hence, by this method it is thinkable to achieve a
scenario in which the magnetic and the hyperfine-interaction become
comparable even for ground state atoms.

We focus on atoms possessing a single active valence electron with
spin $S=\frac{1}{2}$ and a nucleus with spin $I=\frac{3}{2}$ in a
three dimensional magnetic quadrupole field.  In detail we proceed as follows. Section \ref{txt:sec2} contains the
derivation of the underlying Hamiltonian.  Our computational method to calculate the resonance
energies and lifetime is outlined in section \ref{txt:sec4}. Section
\ref{txt:sec5} contains a presentation and discussion of our
results. We conclude with a summary in section \ref{txt:sec6}.

\section{Hamiltonian}\label{txt:sec2}
Taking into account the hyperfine interaction, the Hamiltonian
describing the motion of an atom with mass $M$ and electronic and
nuclear spin $S$ and $I$, respectively, reads in a magnetic
quadrupole field
\begin{equation}\label{ff}
H=\frac{{\bf{p}}{^2}}{2M}+H_{HF}+H_{B}
\end{equation}
where $H_{HF}$ describes the hyperfine interaction between the
outermost single valence electron and the nucleus
\begin{equation}
H_{HF}=A{\bf{I}} \cdot {\bf{S}}.
\end{equation}
Here $A$ is a constant which for an s-electron is given by
\cite{Brandsen}
\begin{equation}
A=\frac{2}{3}
\frac{\mu_{0}}{\hbar^2}g_{e}g_{n}\mu_{B}\mu_{N}|\psi_{s}(0)|^2
\end{equation}
$g_{e}$ and $g_{n}$ are the $g$-factors of the electron and the
nucleus, respectively, and $\psi_{s}(0)$ is the value of the valence
s-electron wave function at the nucleus.
$\mu_{B}=\frac{e\hbar}{2m_{e}}$ and $\mu_{N}=\frac{e\hbar}{2m_{p}}$
are the Bohr magneton for the electron and the proton, respectively.
$H_{B}$ accounts for the interaction of the magnetic moment
of the electron and the nucleus with the magnetic field

\begin{equation}
H_{B}=\mu_{B}g_{e}({\bf{S}}+\alpha {\bf{I}})\cdot {\bf{B}}
\end{equation}
with 
\begin{equation}
\alpha=-\frac{\mu_{N}g_{n}}{\mu_{B}g_{e}}\approx -\frac{m_{e}}{m_{p}}.
\end{equation}
The vector of the three-dimensional quadrupole magnetic
field is given by ${\bf{B}}({\bf{r}})= b (x, y, -2z)$.  By substituting equations (2-5) into equation (1) and performing the scale transformation
\begin{displaymath}
\bar{p}_{i}=(2Mb\mu_{B}g_{e})^{-1/3}p_{i}
\quad\mathrm{and}\quad\bar{x}_{i}=(2Mb\mu_{B}g_{e})^{1/3}x_{i}
\end{displaymath}
we obtain
\begin{eqnarray}
H & = & \frac{1}{2}[{\bf{p}}^2+x(S_{x}+\alpha I_{x})
+ y(S_{y}+\alpha I_{y})\nonumber\\ & & {}-2z(S_{z}+\alpha I_{z})+\beta {\bf{I}} \cdot {\bf{S}}]\label{Hamiltonian}
\end{eqnarray}
with $\beta = \frac{A}{b \mu_{B} g_{e}}$, where the bar have been omitted. The
energy is now measured in units of $\frac{1}{M}(2 M b\mu_{B}
g_{e})^{2/3}$. For convenience and in anticipation of the
forthcoming discussion we transform the Hamiltonian to a spherical
coordinate system, i.e. $(x,y,z)\rightarrow (r,\theta,\phi)$.
Writing the momentum operator explicitly and using atomic units we
obtain
\begin{eqnarray}
H & = & \frac{1}{2} [ -\frac{\partial^2}{\partial
r^2}-\frac{2}{r}\frac{\partial}{\partial
r}+\frac{L^2}{r^2} + r \sin \theta
\cos\varphi(S_{x}+\alpha I{x})\nonumber \\ & & {} + r \sin
\theta\sin\varphi(S_{y}+\alpha I_{y}) - 2r
\cos \theta (S_{z}+\alpha I_{z})\nonumber\\ & & {} + \beta {\bf{I}}
\cdot {\bf{S}}].\label{scaled Hamiltonian}
\end{eqnarray}

Exploring the symmetries of the Hamiltonian we find 16 discrete symmetry operations and a continuous symmetry generated by $J_{z}$, which is the z-component of the total angular momentum of the atom.

The Hamiltonian (\ref{Hamiltonian}) basically differs from the Hamiltonian (4) in ref.\cite{Lesanovsky05} by the additional hyperfine interaction term.  This term does not introduce new symmetries and degeneracies except that all symmetry operations found in ref.\cite{Lesanovsky05} have now to be generalized to the case of the additional presence of the nucleus angular momentum.  For a discussion of degeneracy and symmetries in more details, we refer the reader to ref.\cite{Lesanovsky05}.

\section{Numerical approach}\label{txt:sec4}
The Hamiltonian (\ref{Hamiltonian}) does not support bound states.
Its continuous spectrum is characterized by resonances which are
localized in space (at $t=0$). The time evolution of these states is
given by
\begin{equation}
\psi_{R}(t)=e^{\frac{-iEt}{\hbar}}\psi_{R}(0)
\end{equation}
where $E$ is complex
\begin{equation}
E = \varepsilon - i\frac{\Gamma}{2}.
\end{equation}
Here $\varepsilon$ and $\Gamma$ are the energy and decay width of
the resonance, respectively. Because of the imaginary part
$-i\frac{\Gamma}{2}$ the resonances decay exponentially with a
lifetime of $\tau=\Gamma^{-1}$. In order to calculate the energies
and decay widths of the scattering wave functions we employ the
complex scaling method (see ref.\cite{Moiseyev} and references
therein). The complex scaling transformation only affects the
continuum, while the bound states do not change their positions. The
continuum states are rotated around the corresponding threshold into
the lower half complex energy plane. Eigenvalues belonging to
resonances, once revealed, maintain their positions.

Since resonance states of the complex scaled Hamiltonian are square
integrable the linear variational principle can be applied.
Expanding the wavefunction in a set of basis functions and
calculating the expansion coefficients results in a large-scale
algebraic generalized eigenvalue equation which can be solved by
employing a Krylov space method \cite{Sorensen95}. The basis set must be
chosen in such a way that the exact wave function can be
approximated to a sufficient degree of accuracy by as small as
possible number of functions.  Therefore the form of the basis set
must be adapted to the geometry and the symmetries of the system. We
found a so-called Sturmian basis set of the form
\begin{equation}
\mid n, l, m_{i}, m_{s} \rangle_{m} = R_{n}^{(\zeta)}(r)
Y_{l}^{m-m_{i}-m_{s}}(\theta,\varphi)\chi^{i}(m_{i})\chi^{s}(m_{s})\label{basis}
\end{equation}
suitable for our purposes.  Here the functions $Y_{l}^{m}$
are the spherical harmonics. For fixed $m$ the linear variational
combination of the basis functions $\psi^{m}_{n,l,m_{i},m_{s}}$,
yields, per construction, eigenstates of the Hamiltonian and $J_{z}$
simultaneously. For expanding the radial part we take the
non-orthogonal set of functions
\begin{equation}
R_{n}^{(\zeta)}(r) = e^{-\frac{\zeta r}{2}}L_{n}(\zeta r)
\end{equation}
where $L_{n} (x) $ are the Laguerre polynomials.  The free parameter, $\zeta$,
has the dimension of an inverse length and can be tuned to improve
the convergence behavior in different regions of the energy spectrum.
It should be chosen such that $1/\zeta$ corresponds to the typical length scale
of the states to be approximated.
Using the basis set (\ref{basis}) all matrix elements of the
Hamiltonian(\ref{scaled Hamiltonian}) can be calculated analytically.

\section{Results}\label{txt:sec5}
Let us now discuss the results we obtained while studying an
atom with hyperfine interaction in a magnetic quadrupole field. We
present the resonance energies and decay widths for different values
of the field gradient. The resulting spectrum consist of several
well-separated parts. Concerning the resonance positions one can
distinguish three regimes, each of which reveals individual
characteristics: the weak, the intermediate, and the strong gradient
regimes. In the weak gradient regime, the Zeeman term $H_{B}$ is
very small compared with the hyperfine interaction $H_{HF}$ and only
slightly perturbs the zero-field eigenstates of $H$. In this
case the atom, being primarily in its hyperfine ground state $F=I-\frac{1}{2}$, remains in this manifold and behaves approximately like a neutral particle of spin
$F$ with the $g$-factor
\begin{equation}
g_{f}\simeq g_{e}\frac{F(F+1)-I(I+1)+S(S+1)}{2F(F+1)}.
\end{equation}
This regime is also called the Zeeman regime. In the intermediate
gradient regime the Zeeman and the hyperfine interactions are of the
same order of magnitude, and the atom in the ground state may
represent a significant admixture of different hyperfine states $F$.
Finally in the strong gradient regime the Zeeman term dominates the
hyperfine energies at least for sufficiently large distances from
the coordinate origin.  In this case the spin component of the
electron along the local direction of the magnetic field is almost
conserved and as we will see, the resonance positions in the complex
energy plane are grouped according to different values of its
quantum number $m_{s}$.  This regime is also called the Paschen-Back
Regime.

Knowing the resonance eigenfunctions of the complex-scaled
Hamiltonian one can calculate corresponding expectation values
within the generalized inner c-product \cite{Moiseyev}. The
expectation value which is obtained in this way is in general
complex. The real part represents the average value, whereas the
imaginary part can be interpreted as the uncertainty of our
observable in a measurement when the system is prepared in the
corresponding resonance state \cite{Moiseyev}. We will analyze the
average values of the components of the spins which point along the
local direction of the field as a function of the energy. This
enable us to explain different sets of resonances in each regime. We
also discuss in this section the dependence of the decay width of a
resonance state on its angular momentum as well as on the field
gradient.

\subsection{Resonance Positions in the Zeeman Regime}
\begin{figure*}[h]
$\begin{array}{c}
\includegraphics[height=16cm,width=12cm]{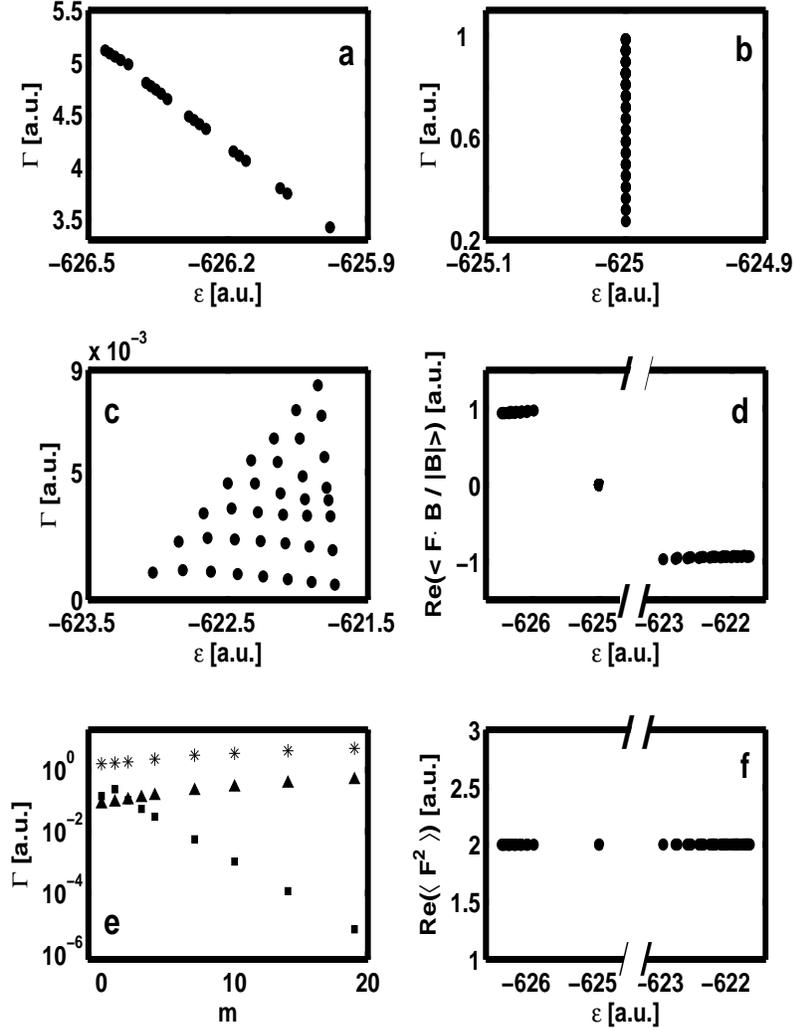}\\
\end{array}$
\caption{Decay width and energies of the resonances for $\beta =
1000$, $m=10$ for (a) $m_{F}=1$, (b) $m_{F}=0$ and (c) $m_{F}=-1$.
(d) Expectation values of the total spin component along the local
direction of the magnetic field for the three sets of resonances.
(e) ($\blacksquare$) The decay width of the energetically lowest
state with $m_{F}=-1$, ($\blacktriangle$) the decay width of the
energetically highest state with $m_{F}=0$, ($\divideontimes$) the
decay width of the energetically highest state with $m_{F}=+1$ as a
function of the quantum number m. (f) The expectation value
of the total spin squared as a function of the energy.} \label{fig1}
\end{figure*}
For large values of $\beta$ ($\beta > 200$) we are in the Zeeman regime. In
Fig. \ref{fig1}(a-c) we present the energies and decay widths for
$\beta = 1000$ and $m=10$, for an atom with nuclear spin $I=\frac{3}{2}$
being in its hyperfine ground state. The resonances are
localized in the negative energy region and their distribution consists of three well-separated parts. They can be
classified according to the expectation value $m_F=Re(\langle \bf{F}\cdot
\bf{B} / |\bf{B}|\rangle)$ which is the projection of the total spin onto the
local direction of the magnetic field. There are long-lived
resonances which can be identified with $m_{F}=-1$, and two sets of
resonances with shorter lifetimes whose spin projections are $m_{F}=
0, +1$. The values of $m_{F}$ are shown in the panel d. For
$m_{F}=-1$ one observes the resonances to be located on lines with
similar slopes covering the area of a right triangle in the
$\epsilon - \Gamma$ plane. For $m_{F}=0$ and $m_{F}=+1$ one
immediately notices that both sets are arranged on lines with an infinity and a negative slope respectively. In Fig.\ref{fig1}(e) we
present the decay width of the energetically lowest state of the
long-lived resonances (the resonances with $m_{F}=-1$) as a function
of the angular momentum $m$.  The decay width decreases
exponentially with increasing value of $m$.  We also present the
decay width of the energetically highest state of the two sets of
short-lived resonances as a function of the quantum number $m$.
Unlike the case $m_{F}=-1$, the decay width of resonances with
$m_{F}=+1$ and $m_{F}=0$ increases with increasing angular momentum.  (for an explanation of this behavior see \cite{Lesanovsky05}).  Fig.\ref{fig1}(f) shows the expectation value of the
squared total spin, $Re(\langle \bf{F}^{2}\rangle)$ as a function of the
energy for the same parameter values.  For all resonance states, the
value is approximately $+2$ which corresponds to $F=1$ i.e. the atom
behaves like a spin-1 particle.

\begin{figure}[h]
$\begin{array}{c}
\includegraphics[height=6cm,width=8cm]{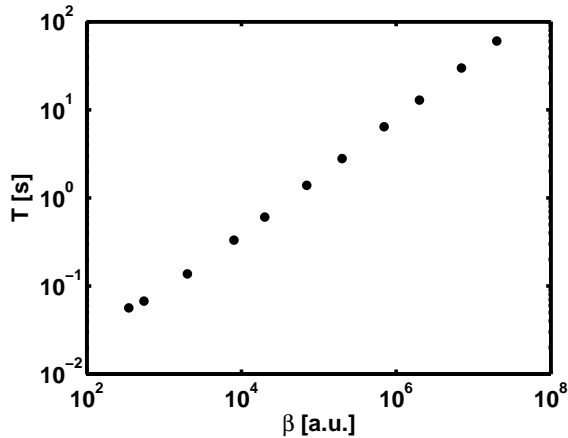}\\
\end{array}$
\caption{The lifetime of the energetically lowest long-lived
resonance state for $m=25$ as a function of the parameter $\beta$.}
\label{fig2}
\end{figure}
In Fig.\ref{fig2} we present the lifetime of the energetically
lowest long-lived resonance state ($m_{F}=-1$) for $m=25$ as a
function of the parameter $\beta$.  The lifetime decreases as the
hyperfine parameter $\beta$ decreases, i.e. the gradient field b
increases. Performing a line fit we find the dependence $T[s]\approx
8.3\times 10^{-4}\beta^{2/3}[a.u.]$.

\subsection{Resonance Positions in the Intermediate Regime}

\begin{figure}[h]
$\begin{array}{c}
\includegraphics[height=7cm,width=9cm]{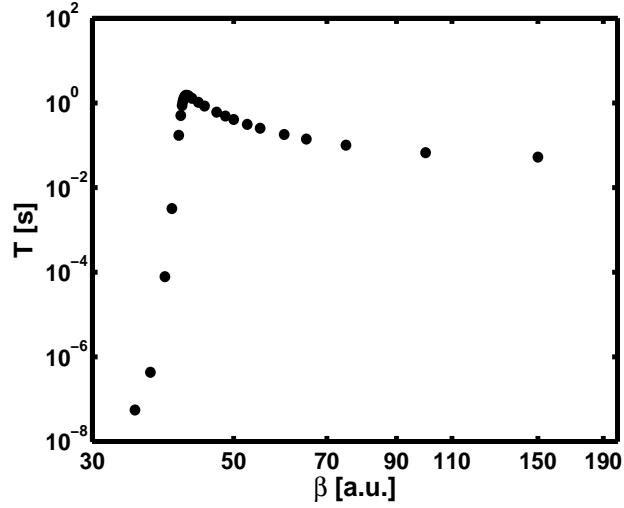}\\
\end{array}$
\caption{The lifetime of the energetically lowest long-lived
resonance state for $m=25$ as a function of the parameter $\beta$.}
\label{fig3}
\end{figure}

Let us now focus on the intermediate regime covering the values
$0.2<\beta <200$, where the Zeeman and hyperfine interactions become
comparable. We observe three different types of behavior. For
$40<\beta <200$, an atom being in the hyperfine ground state,
still behaves approximately like a spin-1 particle.  The resonance spectrum is localized in the negative energy domain and consists of three well-separated parts to which we can assign
the values $m_{F}=-1, 0, +1$. For the two short-lived states, following the same reasoning as above,
the width increases when $m$ increases, while for the long-lived
states it decreases. However in contrast to the Zeeman regime (see Fig.\ref{fig2}), the lifetime of
the long-lived states ($m_{F}=-1$) increases when the hyperfine
parameter $\beta$ decreases, i.e. the field gradient b increases
(see Fig.\ref{fig3}).
For $\beta < 40$ this value decreases very rapidly when $\beta$
decreases. In this case the resonance states which correspond to
higher hyperfine levels $F=2$ are more stable, and are localized in
the positive region of the spectrum.
\begin{figure*}[h]
$\begin{array}{c}
\includegraphics[height=11cm,width=12cm]{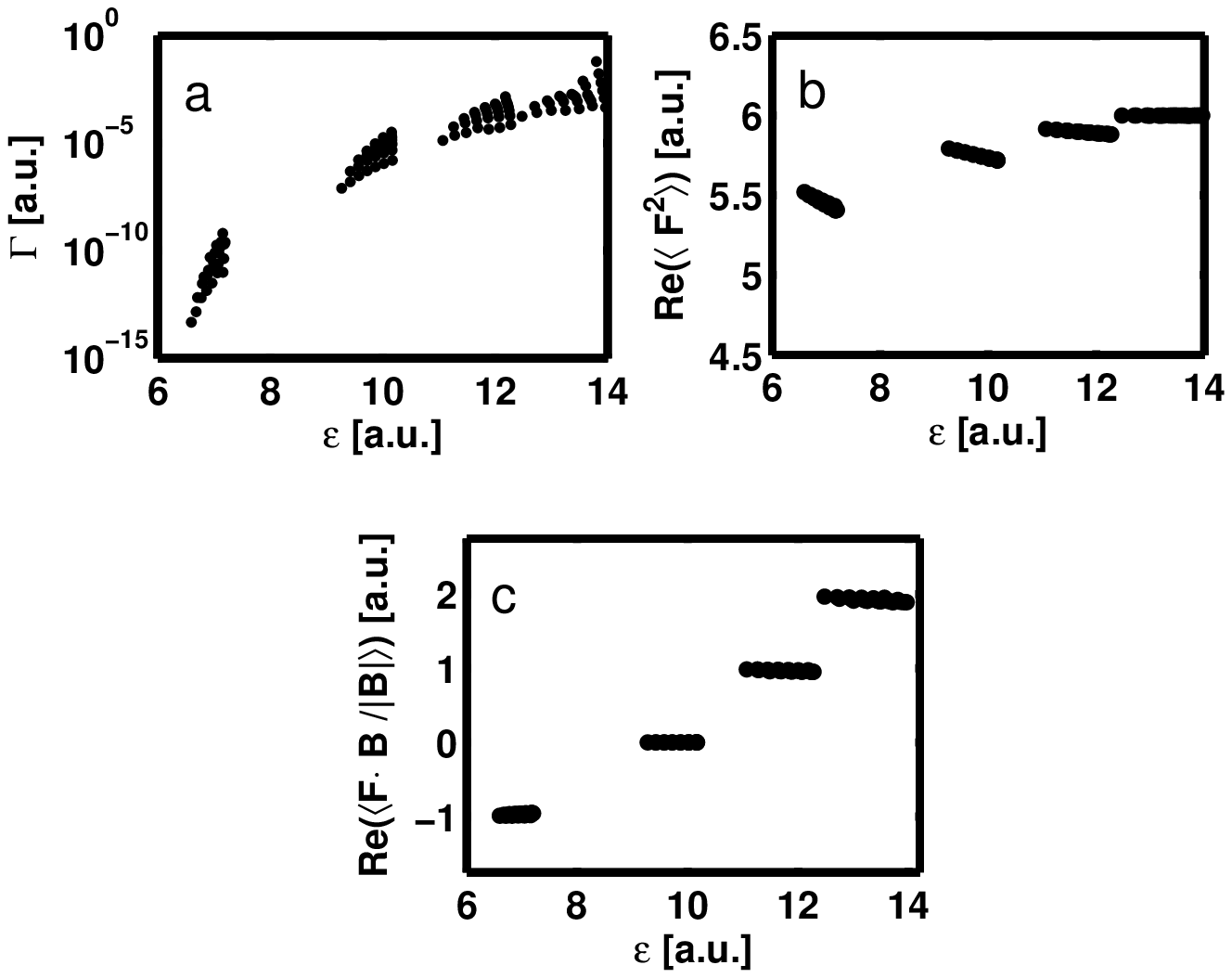}\\
\end{array}$
\caption{(a) Decay width and energies of the long-lived resonances
with higher hyperfine quantum number $F=2$ for $\beta = 19$ and
$m=25$. (b) The expectation value of the total spin squared as a
function of the energy. (c) Expectation values of the total spin
component along the local direction of the magnetic field as a
function of the energy.} \label{fig4}
\end{figure*}
Fig.\ref{fig4}(a) shows the decay widths and energies of the
resonances possessing positive energies for $\beta=19$. We observe
four different curved and triangular shaped regions with distinct
classes of resonances.  In Fig.\ref{fig4}(b) we present the
expectation values of the squared total spin, $Re(\langle
\bf{F}^{2}\rangle)$, of these resonances as a function of the energy. The
values range from approximately $5.5$ to $6.0$, correspond to $F \approx 2.0$. Fig.\ref{fig4}(c) shows the
expectation values of the total spin component along the local
direction of the field for the resonances of Fig.\ref{fig4}(a). The
resonances are divided into four well-separated parts corresponding
to $m_{F}=+2, +1, 0, -1$. Resonances with $m_F = -2$ have
very short lifetime and are not shown. For $\beta < 3$ the
non-conservation of $F$ becomes even more explicit, see
Fig.\ref{fig5}(a) for $\beta=1$. In Fig. \ref{fig5}(b,c) we
present the respective decay widths and energies. The resonances
which are localized in the negative energy region have short
lifetimes and are divided into four subgroups. Each group lies on a
line with negative slope, the four slopes being very similar.
Moreover, the resonances investigated form subgroups on these lines.
The positive energy resonances have much longer lifetimes and cover
an area of approximately triangular shape in an irregular manner,
i.e. no pattern is visible. In Fig.\ref{fig5}(d) the corresponding
expectation values of the electronic spin component along the local
direction of the magnetic field, $Re(\langle \bf{S}\cdot \bf{B}/|\bf{B}|\rangle)$
is shown. For the negative energy resonances this value is
approximately -0.5 indicating that the spin is aligned opposite to
the local direction of the magnetic field, while for positive
energies the spin is parallel to the magnetic field.
Fig.\ref{fig5}(e) shows the expectation values of the nuclear spin
component along the local direction of the field. In the negative
energy domain of the spectrum the pattern divides into four parts
with values $m_{I}=-\frac{3}{2}, -\frac{1}{2}, +\frac{1}{2}, +\frac{3}{2}$,  while in the positive
energy region, the pattern is strongly disturbed and $m_{I}$ is not
conserved. Fig.\ref{fig5}(f) presents the lifetime of the
energetically lowest long-lived state as a function of $\beta$. This
value increases when $\beta$ decreases, i.e. the gradient field b
increases.  Our results show that for the short-lived states, i.e.
the negative energy resonance states, the width increases when $m$
increases, while for the long-lived states, i.e. the positive energy
resonance states, it decreases.
\begin{figure*}[h]
$\begin{array}{c}
\includegraphics[height=16cm,width=12cm]{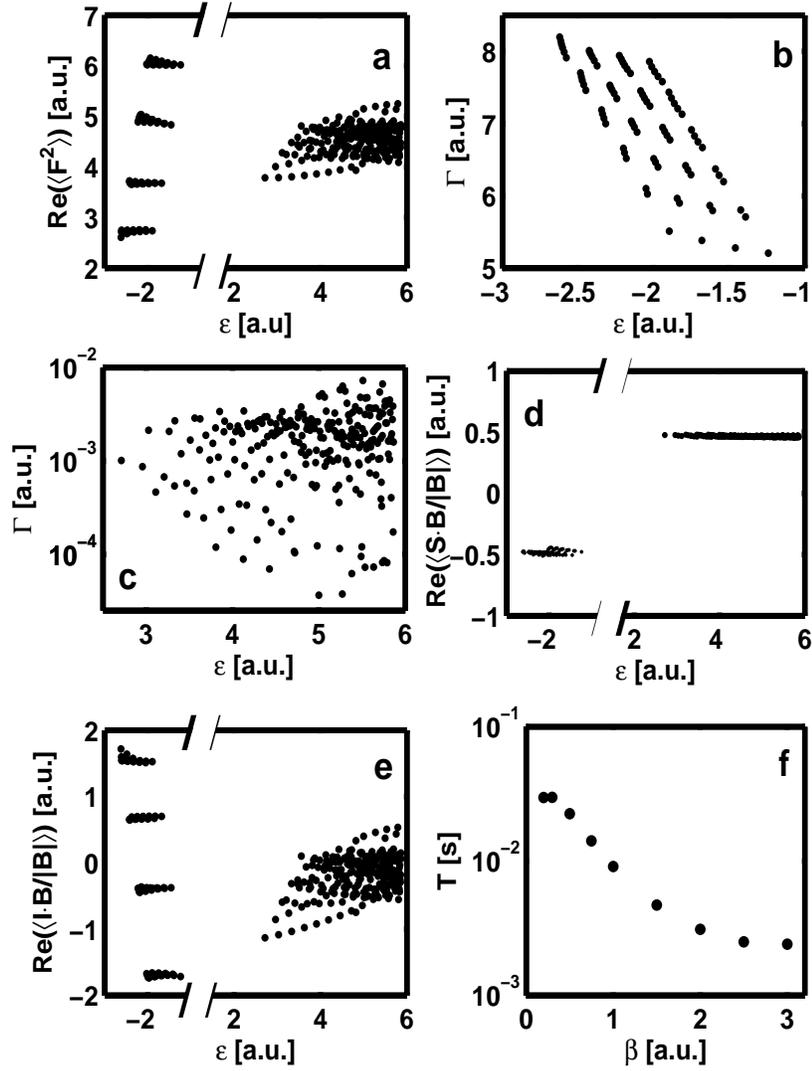}\\
\end{array}$
\caption{(a) The expectation value of the total spin squared as a
function of the energy for $m=10$ and $\beta=1.0$. Decay width and
energies of the resonances for (b) $m_S=-\frac{1}{2}$, (c) $m_S=+\frac{1}{2}$. (d)
Expectation values of the electronic spin component along the local
direction of the magnetic field. (e) Expectation values of the
nuclear spin component along the local direction of the magnetic
field. (f) The lifetime of the energetically lowest resonance state
of the positive energy domain of the spectrum as a function of $\beta$.} \label{fig5}
\end{figure*}

\subsection{Resonance Positions in the Hyperfine Paschen-Back Regime}
\begin{figure*}[h]
$\begin{array}{c}
\includegraphics[height=16cm,width=12cm]{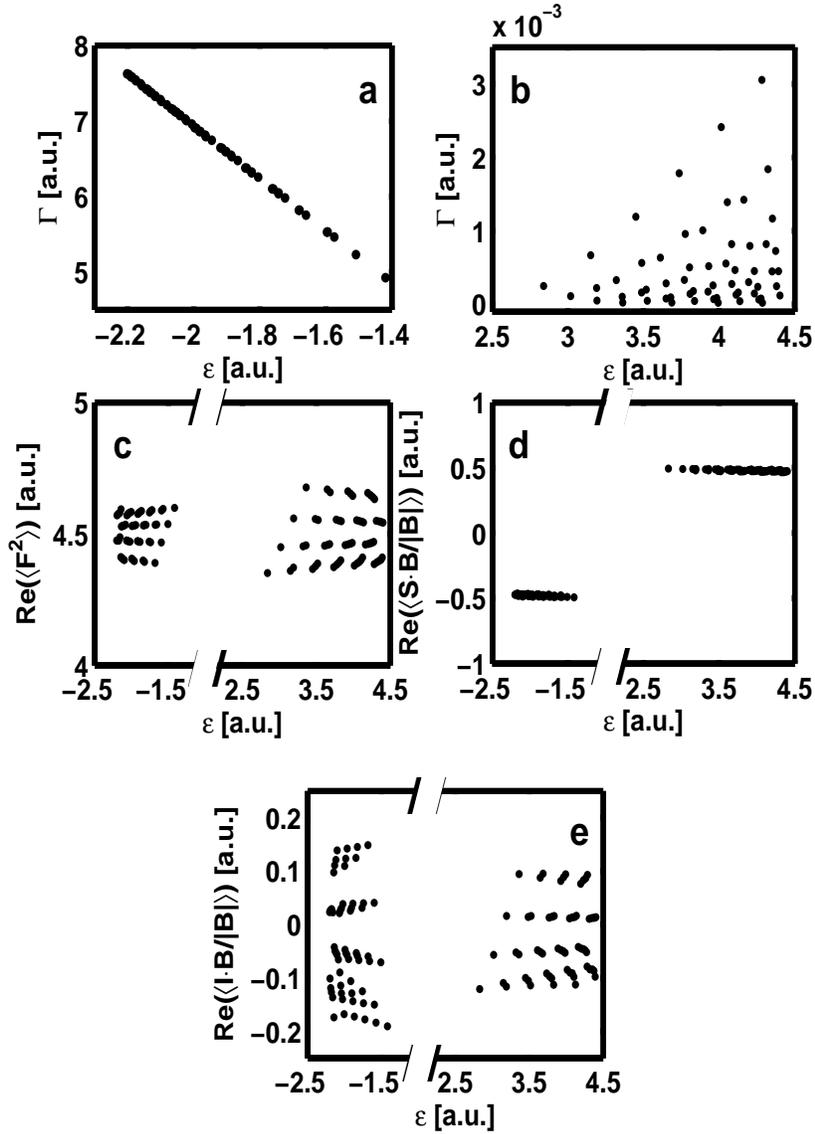}\\
\end{array}$
\caption{Decay width and energies of the resonances for $\beta =
0.1$, $m=10$ for (a) $m_{S}=-\frac{1}{2}$, (b) $m_{S}=+\frac{1}{2}$. (c) The
corresponding expectation value of the total spin squared as a
function of the energy. (d) Expectation values of the electronic
spin component along the local direction of the magnetic field for
the two sets of resonances. (e) Expectation values of the nuclear
spin component along the local direction of the magnetic field.}
\label{fig6}
\end{figure*}

The hyperfine Paschen-Back regime includes $\beta < 0.2$. In case
the Zeeman term dominates the hyperfine energy, it is natural to
decompose the Hamiltonian according to $H=H_{0}+H_{1}$ where
$H_{0}=\frac{1}{2}(\bf{p}^{2}+xS_{x}+yS_{y}-2zS_{z})$ describes the
motion of a spin $\frac{1}{2}$ particle in the magnetic field, and
$H_{1}=\alpha (xI_{x}+yI_{y}-2zI_{z})+\beta \bf{I}\cdot \bf{S}$ perturbs the
eigenstates of $H_{0}$.  The spectrum consists of two parts: again
we have one set of resonances localized in the negative energy
region with short lifetimes and a second set localized in the
positive energy domain possesses much larger lifetimes. In
Fig.\ref{fig6}(a,b) we present the energies and decay widths of
resonances for $\beta =0.1$ and $m=10$. The negative energy
resonances are arranged on lines with a negative slope. The positive
energy resonances cover an area of triangular shape, some of them
being located on straight lines.
Fig.\ref{fig6}(c) shows the corresponding expectation value $Re(\langle
\bf{F}^{2}\rangle )$ of the squared total spin. This value is almost $4.5$ which again indicates that
the total spin quantum number $F$ is not conserved. In
Fig.\ref{fig6}(d) the respective expectation value $Re(\langle \bf{S}\cdot \bf{B}/
|\bf{B}|\rangle)$ of the electronic spin component which points along the local direction of the magnetic field is presented. For the positive energy resonances this value is
approximately +0.5 indicating that the spin is aligned parallel to
the local direction of the magnetic field while for negative energy
resonances it is antiparallel.  The nuclear spin component along the
local field is not conserved, however it is instructive to consider
its average value, $Re(\langle \bf{I}\cdot \bf{B}/|\bf{B}|\rangle)$, as a function
of the energy.  For large values of $m$ (see Fig.\ref{fig6}(e) for
$m=10$) the resonances are grouped into four regular subparts for
both the positive and negative energy domain.

We remark that the decay widths of the positive energy resonances show again
an exponential decaying behavior as a function of the quantum number
$m$ whereas the decay widths of the negative
energy resonance states increases with increasing $m$. The
lifetime of the energetically lowest long-lived resonances for
$m=25$ as a function of $\beta$ can be fitted according to $T [s]
\approx 1.1\times 10^{-1}\beta^{2/3} [a.u.]$.

\section{Summary}\label{txt:sec6}
We have investigated the resonant quantum properties of an atom with
a single active valence electron taking into account its hyperfine
structure ($I=\frac{3}{2}$) in a 3d magnetic quadrupole field. 

We have calculated and analyzed the energies and decay widths of the
resonance states of the Hamiltonian employing the complex scaling
approach and a Sturmian basis set. With respect to the resonance
position one can distinguish essentially three regimes.  In the weak
gradient regime, the Zeeman term is very small compared with the
hyperfine interaction and only slightly perturbs the zero-field
eigenstates of the Hamiltonian.  In this case the atom being primarily in its
hyperfine ground state behaves approximately like a particle of spin
1, and the resonance states are grouped into three well-separated
parts, corresponding to three different directions of the total spin
with respect to the local direction of the field.  The resonances for which the total spin is
antiparallel to the field ($m_F=-1$) possess the largest lifetimes.  As
the total angular momentum of the atom increases, the decay width of
a short-lived resonance state increases while for a long-lived
resonance state it decreases.  Our results for the long-lived resonances show that as the scaled hyperfine parameter $\beta$ increases so does the lifetime.

In the intermediate regime the Zeeman and hyperfine interactions are
comparable. For a sufficiently weak field gradient the atom in the
hyperfine ground state behaves approximately like a spin-1 particle,
and the resonances are arranged somehow similar to the Zeeman regime
but in contrast to the Zeeman regime its long-lived resonance states
become more stable when the gradient field increases and/or $\beta$
decreases.  For stronger field gradients and/or smaller values of
the parameter $\beta$ the lifetime decreases rapidly. In this case
resonances which correspond to the higher hyperfine level, $F=2$,
are more stable.  For even smaller values of $\beta$, the
non-conservation of $F$ becomes remarkable. In this case the
electronic spin component along the local direction of the magnetic
field is almost conserved.  A resonance which is localized in the
negative energy region of the spectrum possesses a short lifetime.
In such a state the electronic spin is aligned opposite to the local
direction of the magnetic field.  The positive energy resonances
possess a much longer lifetime with their electronic spin being
parallel to the magnetic field.  For a
short-lived state the width increases with increasing $m$, while for
a long-lived state it decreases. For the latter our results show that the
lifetime of this state increases with decreasing values of $\beta$.

In the strong gradient regime, the Zeeman term dominates the
hyperfine energy. Here the component of the electronic spin along
the local direction of the field is almost conserved.  The energy
spectrum is arranged into two disconnected parts each of which
contains exclusively resonance states with negative and positive
energies.  The former are of short-lived character and the
electronic spin is antiparallel to the local direction of the
magnetic field. The latter possess much longer lifetimes and the
spin of the electron is parallel to the field.  As the total angular
momentum of the atom increases, the decay width of a negative energy
resonance state increases while for a positive energy resonance
state it decreases.  Similar to the weak gradient
regime the lifetime of a positive energy resonance state, decreases with decreasing $\beta$.

S.S. acknowledges financial support by the Ministry of Science, Research and
Technology of Iran.

\end{document}